\documentclass[twocolumn,twoside,slac_two]{revtex4}
\usepackage{graphicx}
\usepackage{fancyhdr}
\begin{document}

\title{A panchromatic view of relativistic jets in $\gamma$-ray emitting narrow-line Seyfert 1 galaxies}

\author{F. D'Ammando$^{1,2}$, M. Orienti$^1$, J. Finke$^3$, M. Giroletti$^1$, J. Larsson$^4$, on behalf of the {\em Fermi} Large Area Telescope Collaboration}
\affiliation{$^{1}$INAF - Istituto di Radioastronomia, Via Gobetti 101, I-40129 Bologna, Italy}
\affiliation{$^{2}$Dip. di Fisica e Astronomia, Universit\'a di Bologna, Viale Berti Pichat 6/2, I-40127 Bologna, Italy}
\affiliation{$^{3}$U.S. Naval Research Laboratory, 4555 Overlook Ave. SW, Washington, DC 20375-5352, USA} 
\affiliation{$^{4}$KTH, Dep. of Physics, and the Oskar Klein Centre, AlbaNova, SE-106 91 Stockholm, Sweden}

\begin{abstract}
Before the launch of the {\em Fermi} satellite only two classes of Active
Galactic Nuclei (AGN) were known to
generate relativistic jets and thus to emit up to the $\gamma$-ray energy range:
blazars and radio galaxies, both hosted in giant elliptical galaxies. The
first four years of observations by the Large Area Telescope (LAT) on board
{\em Fermi} confirmed that these two populations represent the most numerous identified
sources in the extragalactic $\gamma$-ray sky, but the discovery of variable
$\gamma$-ray emission from 5 radio-loud Narrow-Line Seyfert 1 (NLSy1) galaxies
revealed the presence of a possible emerging third class of AGN with
relativistic jets. Considering that NLSy1 are thought to be hosted in spiral
galaxies, this finding poses intriguing questions about the nature of these
objects, the knowledge of the development of relativistic jets, and the
evolution of radio-loud AGN. In this context, the study of the radio-loud
NLSy1 from radio to $\gamma$-rays has received increasing attention. Here we
discuss the radio-to-$\gamma$-rays properties of the $\gamma$-ray emitting
NLSy1, also in comparison with the blazar scenario.
\end{abstract}

\maketitle

\thispagestyle{fancy}

\section{Introduction}

Relativistic jets are the most extreme expression of the power that can be
generated by a superluminal  black hole (SMBH) in the center of an active
galactic nucleus (AGN), with a large fraction of the power emitted in $\gamma$-rays.
Before the launch of the {\it Fermi} satellite only two classes of AGN were
known to generate these structures and thus to emit
up to the $\gamma$-ray band: blazars and radio galaxies, both hosted
in giant elliptical galaxies \cite{blandford78}. The
first 4 years of observation by the Large Area Telescope (LAT) on board {\em
  Fermi} confirmed that the extragalactic $\gamma$-ray sky is dominated by
radio-loud AGN, being mostly blazars and some radio galaxies \cite{acero15}. However, the
discovery by {\em Fermi}-LAT of variable $\gamma$-ray emission from a few
radio-loud narrow-line Seyfert 1 (NLSy1) galaxies revealed the presence of a possible
third class of AGN with relativistic jets \cite{abdo09}. 

NLSy1 are a class of AGN identified by \cite{osterbrock85}
and characterized by their optical properties: narrow permitted lines (FWHM
(H$\beta$) $<$ 2000 km s$^{-1}$),
[OIII]/H$\beta$ $<$ 3, and a bump due to Fe II (e.g., \cite{pogge00}). They also exhibit strong X-ray variability, steep X-ray
spectra, relatively high luminosity, and substantial soft X-ray excess (e.g., \cite{grupe10}). These characteristics point to systems with smaller
masses of the central black hole (10$^6$--10$^8$ M$_\odot$) than blazars and radio galaxies, and higher accretion rates (close to or above the Eddington limit). NLSy1 are generally radio-quiet (radio-loudness $R<$ 10), with only
a small fraction ($<$ 7$\%$; \cite{komossa06}) classified as radio-loud. Objects with high
values of radio-loudness ($R>$ 100) are even more sparse ($\sim$2.5\%), while
generally $\sim$15$\%$ of quasars are radio-loud. Considering also that NLSy1
are thought to be hosted in spiral galaxies, their detection in $\gamma$-rays poses intriguing questions about the
nature of these sources, the production of relativistic jets, the
mechanisms of high-energy emission, and the cosmological evolution of radio-loud AGN.

\section{The $\gamma$-ray view of NLSy1}

\begin{figure}[t]
\centering
\includegraphics[width=65mm]{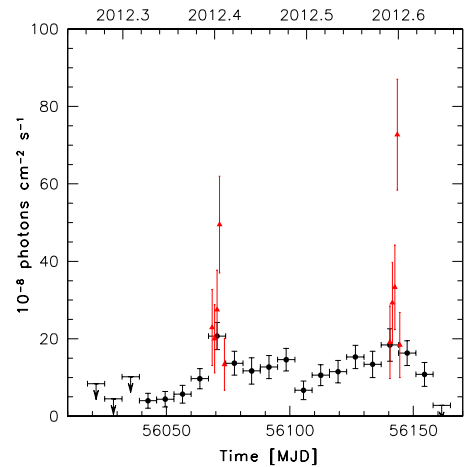}
\caption{{\em Fermi}-LAT (E $>$ 100 MeV) light curve of SBS 0846$+$513
  obtained during 2012 April 1 - August 28 with 7-day (black circles) and
  1-day (red triangles) time bins. Arrows refer to 2-$\sigma$ upper
  limits. Upper limits are computed when TS $<$10. Adapted from \cite{dammando13e}.} \label{Fig1}
\end{figure}

Five radio-loud NLSy1 galaxies have been detected at high significance
by {\em Fermi}-LAT so far: 1H 0323$+$342, SBS 0846$+$513, PMN J0948$+$0022, PKS
1502$+$036, and PKS 2004$-$447 \cite{abdo09, dammando12, acero15}, with a
redshift between 0.061 and 0.585. The average apparent isotropic luminosity of
these sources in the 0.1--100 GeV energy band is between
10$^{44}$ erg s$^{-1}$ and 10$^{47}$ erg s$^{-1}$, a range of values typical
of blazars \cite{dammando13b}. This may be an indication of a small viewing angle with respect to the jet axis and thus a high beaming
factor for the $\gamma$-ray emission, similarly to blazars. 
Several strong $\gamma$-ray flares were observed
from SBS 0846$+$513 (Fig.~\ref{Fig1}) and PMN J0948$+$0022, reaching a peak apparent isotropic $\gamma$-ray
luminosity of $\sim$10$^{48}$ erg s$^{-1}$, comparable to that of the bright
FSRQ \cite{foschini11,dammando13e,dammando15}. In particular, SBS 0846$+$513 and PMN J0948$+$0022 showed a $\gamma$-ray flaring activity combined with a moderate spectral evolution \cite{dammando12,foschini11}, a behaviour that was already observed in bright
flat spectrum radio quasars (FSRQ) and low-synchrotron-peaked BL Lacs \cite{abdo10}. Variability and spectral
properties of these two NLSy1s in $\gamma$-rays indicate a blazar-like behaviour. An intense $\gamma$-ray flaring activity was observed by
LAT also from 1H 0323$+$342 \cite{carpenter13}. This is another indication that radio-loud NLSy1 are able to host relativistic jets as powerful as those in blazars.

\section{X-ray properties}

The X-ray spectra of NLSy1 are usually characterized by a steep photon index
($\Gamma_{\rm\,X}$ $>$ 2, \cite{grupe10}). On the contrary, a relatively hard
X-ray spectrum was detected in the {\em Swift}/XRT
observations of SBS 0846+513 \cite{dammando12,dammando13d}, PMN J0948$+$0022
\cite{foschini11, dammando14, dammando15,foschini12}, 1H 0323$+$342 \cite{dammando13c}, and PKS
1502$+$036 \cite{dammando13a}. This suggests a significant contribution of inverse Compton radiation from a relativistic jet,
similar to what is found for FSRQ. 

\begin{figure}[t]
\centering
\includegraphics[width=50mm, angle=-90]{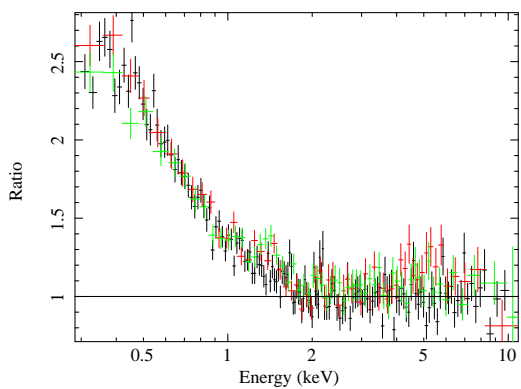}
\caption{{\em XMM-Newton} EPIC pn (black), MOS1 (red) and MOS2 (green) data of
  PMN J0948$+$0022 shown as a ratio to a power law with $\Gamma$ =
  1.48. Adapted from \cite{dammando14}.} \label{Fig2}
\end{figure}

The high quality {\em XMM-Newton} observation of PMN
J0948$+$0022 performed in 2011 May allowed us to study in
detail its X-ray spectrum, as reported in \cite{dammando14}. The spectral
modelling of the {\em XMM-Newton} data of PMN
J0948$+$0022 shows that emission from the jet most likely
dominates the spectrum above $\sim$2 keV, while a soft X-ray
excess is evident below $\sim$2 keV (Fig.~\ref{Fig2}). The origin of the soft X-ray excess is still an open
issue both in radio-quiet and radio-loud AGN (e.g., \cite{gierlinski04}). Such a Seyfert component is a
typical feature in the X-ray spectra of radio-quiet NLSy1,
but it is quite unusual in jet-dominated AGN, even if not
unique (e.g., PKS 1510-089; \cite{kataoka08}). In the case of PMN
J0948$+$0022, the statistics did not allow us to distinguish between different
models for the soft X-ray emission. Models where the soft emission is partly
produced by blurred reflection, or Comptonisation of the
thermal disc emission, or simply a steep power-law, all provide
good fits to the data. A multicolor thermal disc emission also
gives a comparable fit, but the temperature is too high (kT = 0.18
keV) and is incompatible with a standard Shakura \& Sunyaev accretion disc \cite{dammando14}.

\begin{figure}[b]
\centering
\includegraphics[width=65mm]{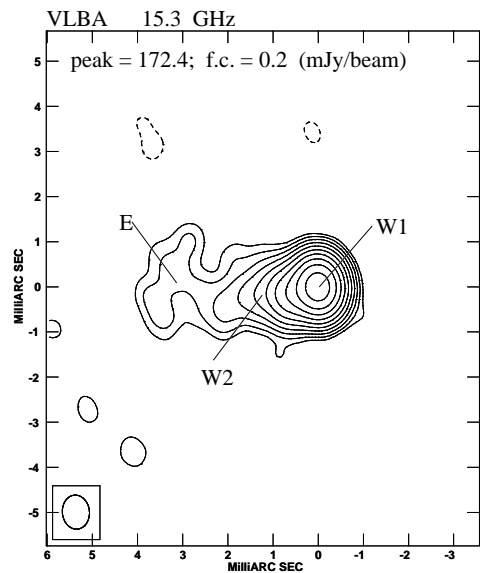}
\caption{15.3 GHz VLBA image of SBS 0846$+$513. On the image we provide the
  peak flux density, in mJy/beam, and the first contour intensity (f.c., in
  mJy/beam) that corresponds to three times the noise measured on the image
  plane. Contour levels increase by a factor of 2. The beam is plotted on the
  bottom left corner of the image. Component W1 is the core region, W2 is a
  knot, and E is the jet structure. Adapted from \cite{dammando13d}.} \label{Fig3}
\end{figure}

\section{Radio properties}

\begin{figure}[t]
\centering
\includegraphics[width=75mm]{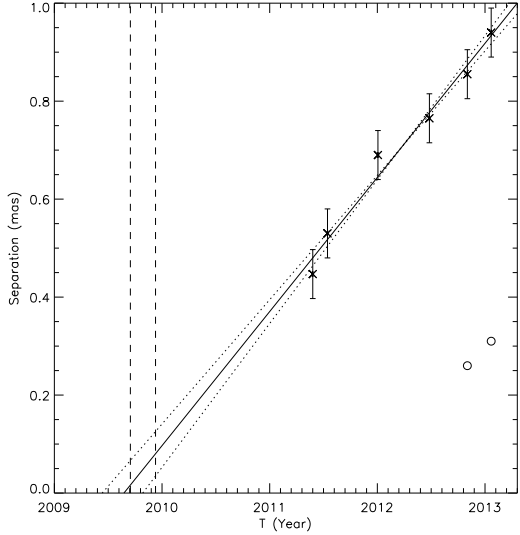}
\caption{The separation between the core component of SBS 0846$+$513 and the knot ejected in 2009 as a function of time. The
solid line represents the regression fit to the 15 GHz VLBA MOJAVE data, while the dotted lines represent the
uncertainties from the fit parameters. Dashed lines indicate the beginning and
the peak of the radio flare. Adapted from \cite{dammando13e}.} \label{Fig4}
\end{figure}

On pc scale a core-jet structure was observed for SBS 0846+513
\cite{dammando12} (Fig.~\ref{Fig3}), PKS 2004$-$447 \cite{orienti12}, PKS 1502$+$036
\cite{dammando13a}, and PMN J0948$+$0022 \cite{giroletti11, dammando14},
although the jet in the two latter sources is significantly fainter than that
observed in the former two sources. The analysis of the 6-epoch data set of SBS 0846$+$513 collected by the MOJAVE programme during 2011-2013
indicates that a superluminal jet component is moving away from the core with an apparent
angular velocity of (0.27$\pm$0.02) mas yr$^{-1}$ (Fig.~\ref{Fig4}), corresponding to
(9.3$\pm$0.6)$c$ \cite{dammando13d}. This apparent superluminal velocity indicates the presence of boosting effects
for the jet of SBS 0846$+$513. On the contrary, VLBA observations did not detect apparent
superluminal motion at 15 GHz for PKS 1502$+$036 during 2002$-$2012, although the radio
spectral variability and the one-sided jet-like structure seem to require the presence of
boosting effects in a relativistic jet \cite{dammando13a}. 

Strong radio variability was observed at 15 GHz during the monitoring of the OVRO 40-m telescope of PMN
J0948+0022 \cite{dammando14,dammando15}, PKS 1502+036 \cite{dammando13a}, and SBS 0846+513 \cite{dammando12,dammando13d}. An inferred variability brightness
temperature of 2.5$\times$10$^{13}$~K, 1.1$\times$10$^{14}$~K, and 3.4$\times$10$^{11}$~K was obtained for PKS 1502$+$036, SBS 0846$+$513, and
PMN J0948$+$0022, respectively. These values are larger than the brightness temperature derived for the
Compton catastrophe \cite{readhead94}, suggesting that the radio emission of the jet is Doppler boosted. On the
other hand, a high apparent brightness temperature of 10$^{13}$ K, comparable to that of the $\gamma$-ray
NLSy1, was observed for TXS 1546$+$353. However, no $\gamma$-ray
emission has been detected from this source, so far \cite{orienti15}. Moreover,
an intensive monitoring of these $\gamma$-ray NLSy1 from 2.6 GHz to 142 GHz
with the Effelsberg 100-m and IRAM 30-m telescopes showed, in addition to an
intensive variability, spectral evolution across the different bands
following evolutionary paths explained by travelling shocks, typical
characteristics seen in blazars \cite{angelakis15}.

\section{Multifrequency variability and SED modelling}

\begin{figure}[t]
\centering
\includegraphics[width=75mm]{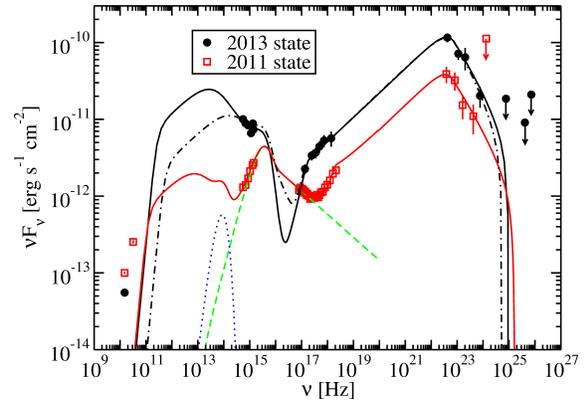}
\caption{SED and models for the 2013 and 2011 activity states of PMN J0948$+$0022. The filled circles are the data
from the 2013 flaring state, and the open squares are the data from the 2011 intermediate state taken from \cite{dammando14}. The dashed curve indicates the disc and coronal emission, and the dotted line indicates the thermal dust
emission. Solid lines represent models consistent with scattering dust torus radiation, while the dashed-dotted curve
represents a model consistent with the scattering of BLR radiation. Arrows refer to 2$\sigma$ upper limits on the source flux.
The VERITAS upper limits are corrected for EBL absorption using the model of
\cite{finke10}. Adapted from \cite{dammando15}.} \label{Fig5}
\end{figure}

The first spectral energy distributions (SED) collected for the four NLSy1s detected in the first year of {\em Fermi} operation showed
clear similarities with blazars: a double-humped shape with a first peak in the IR/optical band
due to synchrotron emission, a second peak in the MeV/GeV band likely due to inverse
Compton emission, and an additional component related to the accretion disc in UV for three of the four sources. The physical
parameters of these NLSy1 are blazar-like, and the jet power is in the
average range of blazars \cite{abdo09}.

For PMN J0948$+$0022 we compared the broad-band SED of the 2013 flaring activity state with that from an
intermediate activity state observed in 2011 (Fig.~\ref{Fig5}). Contrary to what was observed
for some FSRQ (e.g., PKS 0537$-$441; \cite{dammando13e}) the SED of the two
activity states, modelled as synchrotron emission and as an external Compton scattering of seed photons from a dust torus, could not be
modelled by changing only the electron distribution parameters. A higher magnetic field is needed
for the high activity state, consistent with the modelling of different
activity states of PKS 0208$-$512 \cite{chatterjee13}. We also modelled the
2013 flaring state assuming Compton-scattering of broad line region (BLR) line radiation. The model reproduces
the data as well as the scattering of the IR torus photons. However, we note
that the BLR
scattering model requires magnetic fields which are far from equipartition.

We also compared the SED of SBS 0846$+$513 during the flaring state in 2012 May with that of a
quiescent state. Similar to PMN J0948$+$0022, the SED of the two different activity states, modelled by an external Compton
component of seed photons from a dust torus, could be fitted by changing the electron
distribution parameters as well as the magnetic field \cite{dammando13d}. A
significant shift of the synchrotron peak to higher frequencies was observed during the 2012 May
flaring episode, similar to FSRQ (e.g., PKS 1510$-$089; \cite{dammando11}).
Contrary to what is observed in PMN J0948$+$0022, no significant evidence of thermal emission
from the accretion disc has been observed in SBS 0846$+$513.

\begin{figure}[t]
\centering
\includegraphics[width=85mm]{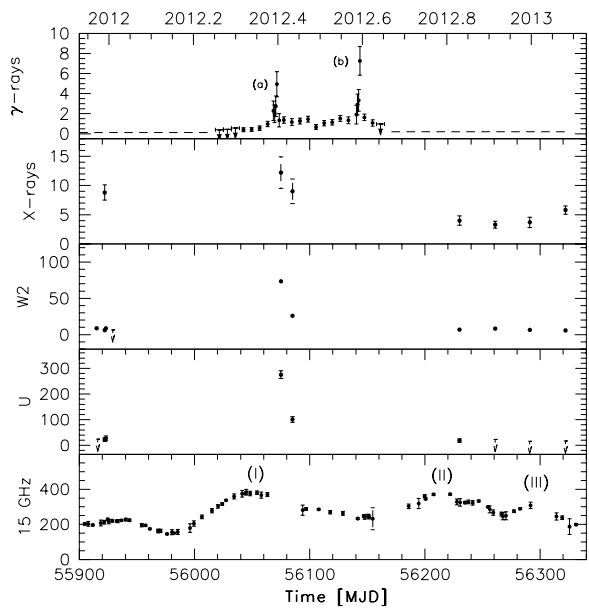}
\caption{Multifrequency light curve for SBS 0846$+$513 during 2011 December -
  2013 January. The data sets were collected (from top to bottom) by {\em
    Fermi}-LAT ($\gamma$-rays, 0.1$-$100 GeV; in units of 10$^{-7}$ ph cm$^{-2}$
  s$^{-1}$), {\em Swift}-XRT (0.3$-$10 keV; in units of 10$^{-13}$ erg cm$^{-2}$
  s$^{-1}$), {\em Swift}-UVOT (w2 and u bands; in units of $\mu$Jy) and OVRO
  (15 GHz; in units of mJy). Arrows refer to 3$\sigma$ upper limits on the
  source flux densities for the w2 and u bands, and to 2$\sigma$ upper limits
  on the source fluxes for the $\gamma$-ray light curve. Adapted from \cite{dammando13d}.} \label{Fig6}
\end{figure}

A complex connection between the radio and $\gamma$-ray emission was observed
for SBS 0846$+$513 and PMN J0948$+$0022, where
$\gamma$-ray and radio flares have not a similar behaviour, as discussed in
detail in \cite{dammando13d,dammando14,foschini12}. Optical intraday
variability has been reported for PMN J0948$+$0022 by \cite{liu10, maune13,
  itoh13}, sometimes associated with a significant increase of the optical
polarisation percentage, indicating
a relativistic jet as the most likely origin for the optical emission in this object.  

At Very High Energy (VHE; E $>$ 100 GeV), VERITAS observations of PMN J0948$+$0022 were carried out during 2013 January 6--13, after the $\gamma$-ray flare observed by {\em Fermi}-LAT on 2013 January 1. These observations resulted in an upper limit of F$_{> 0.2\rm\,TeV}$ $<$ 4$\times$10$^{-12}$ ph cm$^{-2}$ s$^{-1}$ \cite{dammando15}. The lack of detection at VHE could be due to different reasons: 1) The
distance of the source ($z$ = 0.5846) is relatively large and most of the GeV/TeV emission
may be absorbed due to pair production from $\gamma$-ray photons of the
source and the infrared photons from the extragalactic background
light (EBL). However, we must note that the most distant FSRQ detected at VHE
up to now, 3C 279 \cite{albert08}  is at a comparable
distance. 2) The VERITAS observations were carried out a few days after the peak
of the $\gamma$-ray activity, thus covering only the last part of the MeV/GeV
flare. 3) Considering the similarities with FSRQ, a BLR should be present in these NLSy1. The presence of a BLR
could produce a spectral break due to pair production, suppressing the flux beyond a few GeV and preventing a VHE detection. However, the detection at VHE of the
FSRQ 3C 279, PKS 1510$-$089 \cite{abramowski13,aleksic14}, and 4C $+$21.35 \cite{aleksic11} have
shown that the spectrum of some FSRQ extends to VHE energies during some
flares, indicating that the $\gamma$-rays may be produced outside the BLR during those high-activity
periods. The same scenario may apply to the $\gamma$-ray emitting NLSy1.

\section{Radio-loudness, host galaxy, and jet formation}

The mechanism at work for producing a relativistic jet is not clear. In
particular, the physical parameters that drive the jet formation is still under
debate. By considering that NLSy1 are thought to be hosted in spiral
galaxies (e.g., \cite{deo06}) the presence of a relativistic
jet in these sources seems to be in contrast to the paradigm that such
structures could be produced only in elliptical galaxies (e.g., \cite{marscher09}). The most powerful jets are
found in luminous elliptical galaxies with very massive central BH and low accretion rates (e.g., \cite{mclure04,sikora07}). This was interpreted as
an indirect evidence that a high spin is required for the jet production, since
at least one major merger seems to be necessary to spin up the SMBH. At
the same time, low accretion rates, which are associated with geometrically
thick advection dominated accretion flows, may be important in jet formation
by creating large-scale poloidal magnetic fields \cite{sikora13}. 
 
Therefore one of the most surprising facts related to the discovery of NLSy1 in $\gamma$-rays is the
development of a relativistic jet in objects with a relatively small BH mass
of 10$^{7}$-10$^{8}$ M$_\odot$. However, it is worth noting that the mass
estimation of the BH in these sources may have large uncertainties due to the
effect of radiation pressure \cite{marconi08} and the possible disc-like
structure of their BLR \cite{decarli08}. \cite{calderone13} modelling the optical/UV data of some radio-loud NLSy1 with a Shakura
\& Sunyaev disc spectrum have estimated higher BH masses than those reported
in the past, for PMN J0948$+$0022 and PKS 1502$+$036 comparable to the
values estimated for blazars.
This may solve the problem of the minimum BH mass predicted in different scenarios of relativistic jet formation and development, but introduces a new problem. How is
it possible to have such a large BH mass in a class of AGN usually hosted in
spiral galaxies? Only very sparse observations of the host galaxy of
radio-loud NLSy1 are available up to now. Among the NLSy1 detected by LAT
only for the closest one, 1H 0323+342, the host galaxy was clearly detected,
suggesting two possible scenarios: the spiral arms of the host galaxy
\cite{zhou07} or the residual of a galaxy merger \cite{anton08, tavares14}. Therefore the possibility that the production of relativistic jets in these objects could be due to strong merger activity, unusual in disc/spiral galaxies, cannot be ruled out.

The accretion rate (thus the mass) and the spin of the BH seem to
be related to the host galaxy, leading to the hypothesis that relativistic jets
can form only in elliptical galaxies \cite{marscher09,bottcher02}. We
noted that the BH masses of radio-loud NLSy1 are generally larger than those in the whole sample of NLSy1 (M$_{\rm BH}$
$\approx$(2--10)$\times$10$^{7}$ M$_\odot$; \cite{komossa06, yuan08}), even if still small if compared to radio-loud quasars. The larger BH masses of radio-loud NLSy1 with
respect to radio-quiet NLSy1 may be related to prolonged accretion episodes that can spin-up the BHs. In
this context, the small fraction of radio-loud NLSy1 with respect to radio-loud quasars could be
an indication that not in all the former the high-accretion regime lasted
long enough to spin-up the central BH \cite{sikora09}.

\begin{acknowledgments}

The {\em Fermi} LAT Collaboration acknowledges support from a number of agencies and institutes for
both the development and the operation of the LAT as well as
scientific data analysis.  These include NASA and DOE in the United
States, CEA/Irfu and IN2P3/CNRS in France, ASI and INFN in
Italy, MEXT, KEK, and JAXA in Japan, and the
K.~A.~Wallenberg Foundation, the Swedish Research Council and the National Space Board in Sweden. Additional support from INAF in Italy
and CNES in France for science analysis during the operations phase is also
gratefully acknowledged. 

We thank the {\em Swift} team for making these observations possible, the
duty scientists, and science planners. This
research made use of data from MOJAVE database that is maintained by the
MOJAVE team (Lister et al. 2009, AJ, 137, 3718).
The OVRO 40-m monitoring program
is supported in part by NASA grants NNX08AW31G and NNX11A043G, and NSF grants AST-0808050 
and AST-1109911. This work is partly based on observations obtained
with {\em XMM-Newton}, an ESA science mission with instrument and contributions
directly funded by ESA Member States and the USA (NASA).
Part of this work was done with the contribution of
the Italian Ministry of Foreign Affair. FD, MO, MG acknowledge financial
contribution from grant PRIN-INAF-2011.

\end{acknowledgments}

\end{document}